\documentclass[%
aip,
amsmath,amssymb,
reprint,%
]{revtex4-1}

\usepackage{graphicx}% Include figure files
\usepackage{dcolumn}% Align table columns on decimal point
\usepackage{bm}% bold math

\usepackage[utf8]{inputenc}
\usepackage[T1]{fontenc}
\usepackage{mathptmx}

\usepackage{svg}
\usepackage{graphicx}
\usepackage{amsmath}
\usepackage{amssymb}
\usepackage{color}
\usepackage[normalem]{ulem}

\begin{document}
	
	\preprint{AIP/123-QED}
	
	\title[Networks of Bistable Units]{
	Bifurcations of Clusters and Collective Oscillations in Networks of Bistable Units}
	% Force line breaks with \\
	
	\author{Munir Salman}
	\email{munir.salman@tum.de}
	\affiliation{ 
		\mbox{Physics Department, Technical University of Munich, James-Franck-Str. 1, 85748 Garching, Germany}%\\This line break forced with \textbackslash\textbackslash
	}
	
	\author{Christian Bick}
	\affiliation{\mbox
		{Institute for Advanced Study, Technical University of Munich,
			Lichtenbergstr.~2, 85748 Garching b.~M\"unchen, Germany}
	}
	\affiliation{\mbox{Department of Mathematics, Vrije Universiteit Amsterdam, 
	        De Boelelaan 1111, Amsterdam, the Netherlands}
	}			
	\affiliation{\mbox{Centre for Systems Dynamics and Control and Department of
			Mathematics, University of Exeter, Exeter EX4~4QF, UK}
	}

	\author{Katharina Krischer}
	\email{krischer@tum.de}
	\affiliation{ 
		\mbox{Physics Department, Technical University of Munich, James-Franck-Str. 1, 85748 Garching, Germany}%\\This line break forced with \textbackslash\textbackslash
	}

	\date{\today}% It is always \today, today,
	
	\begin{abstract}
		We investigate dynamics and bifurcations in a mathematical model that captures electrochemical experiments on arrays of microelectrodes. In isolation, each individual microelectrode is described by a one-dimensional unit with a bistable current-potential response. When an array of such electrodes is coupled by controlling the total electric current, the common electric potential of all electrodes oscillates in some interval of the current. These coupling-induced collective oscillations of bistable one-dimensional units are captured by the model. Moreover, any equilibrium is contained in a cluster subspace, where the electrodes take at most three distinct states. We systematically analyze the dynamics and bifurcations of the model equations: We consider the dynamics on cluster subspaces of successively increasing dimension and analyze the bifurcations occurring therein. Most importantly, the system exhibits an equivariant transcritical bifurcation of limit cycles. From this bifurcation, several limit cycles branch, one of which is stable for arbitrarily many bistable units.
	\end{abstract}
	
	\maketitle

\newcommand{\dt}{\dot}
\newcommand{\ddt}{{\frac{\mathrm{d}}{\mathrm{d}t}}}
\newcommand{\y}{y}
\newcommand{\cA}{\mathcal{A}}
\newcommand{\redbf}[1]{\textcolor{blue}{\textbf{#1}}}
\newcommand{\hl}[1]{#1}

\begin{quotation}
  The interaction between simple individual units can lead to the spontaneous occurrence of spatial- or/and temporal patterns. A lot of focus has been on the collective dynamics of coupled individual (microscopic) oscillators. However, in some important experimental contexts, the individual units are one-dimensional and bistable rather than oscillatory. Examples include phase transition cathodes in Li-ion batteries and the electrocatalytic oxidation of CO-on-Pt-nanoparticles or arrays of Pt-microelectrodes. For the microelectrode array, one can observe that fixing the total current through the electrodes leads to clustered dynamics of two or three clusters where all units within one cluster take on the same state. Moreover, such clustered states can exhibit sustained collective oscillations. These dynamic features are captured in a simple mathematical model~\cite{salman2020collective}. Here we elucidate the dynamics and bifurcations that arise in the model equations as parameter are varied. We exploit the fact that in a stationary state, each of the~$N$ bistable units takes on only one of three different values and reduce the $N$-dimensional system to two or three degrees of freedom that describe the clustered dynamics. This allows us to perform a bifurcation analysis of the steady states and limit cycles, which arise in Hopf bifurcations. Furthermore, using knowledge of the stability of the clusters in the full system~\cite{salman2020collective}, we can conclude that
  stable oscillations 
  also exist in thermodynamically large systems with~$10^{10}$ individual bistable units, 
  for example, in electrocatalytical systems composed of nanoparticular catalyst particles on some support.
  
\end{quotation}

%--------------------------------------------------------
\section{Introduction}

Networks of interacting bistable units are an important class of dynamical systems that are relevant in diverse physical contexts.
While coupled units with oscillatory or chaotic dynamics have been widely considered (see, e.g., the textbook Ref.~\onlinecite{pikovsky2003synchronization}), 
multi-component systems in which each unit has a single degree of freedom have received much less attention.
Zanette~\cite{zanette1997dynamics} considered globally coupled bistable elements which interacted linearly through their common mean. The global coupling was diffusive in the sense that it vanished for uniform states. For random initial condition and a sufficiently large coupling strength, the global interaction led to a coherent motion of the entire ensemble towards one of the two stable states. 
The impact of additive noise to the bistable units in related network configurations was analyzed in Refs.~\onlinecite{pototsky2009synchronization,Ashwin2017a,atsumi2013phase,pototsky2008hysteresis}, while Ref.~\onlinecite{yamazaki2011collective,rungta2017network, rungta2018identifying} considered the collective response of bistable units that 
were connected through different coupling topologies.
A recent study also examined the transition from a local to a global coupling topology, focusing on the mathematical perspective~\cite{TianIMA2021}.

Kouvaris \emph{et al.}~\cite{kouvaris2013feedback} investigated the effect of a global feedback on the dynamics of networks of diffusively coupled bistable units. The global feedback was chosen such that it altered the excitation threshold, i.e., the position of the saddle point, of the local one-dimensional reaction dynamics. In this case, localized stationary activation patterns form, the size of which can be adjusted by varying the feedback strength. The impact of a global time-delayed feedback and additive noise on an ensemble of bistable units was examined in Refs.~\onlinecite{huber2003dynamics,pototsky2009synchronization}, {while in Ref.~\onlinecite{kohar2013verification} the impact of heterogeneities on the synchronization state of globally coupled bistable electronic circuits is discussed.} 

Here, we consider bistable units that interact through a macroscopic observable which is forced to take a fixed value as a global constraint. This observable can be the sum of the intrinsic state variables of the individual units, or it can be the sum of a function of the individual state variables. The global constraint is enforced by allowing a parameter that controls the state of an individual unit on the equilibrium branch to adapt; thus the global constraint can be seen as a bifurcation parameter of the coupled system. 
Indeed, such coupling between bistable units is realized in many physical systems: These include Li--ion batteries with phase-transition cathodes~\cite{dreyer2010thermodynamic,dreyer2011behavior} and bistable electrochemical reactions, such as the CO oxidation on an array of Pt electrodes~\cite{crespo2013cooperative,crespo2014sequential,bozdech2018oscillations}. In case of Li--ion batteries, the cathodes consist of billions of nano-particles that can be considered as bistable units. Each nanoparticle can be in a Li-rich or a Li-poor state depending on the chemical potential or equivalently the voltage. Thus, the battery can be seen as interacting bistable units coupled through a global constraint: When charging or discharging the battery slowly, a constraint is set to the time evolution of the total charge, while the voltage adjusts accordingly. Similarly, the CO-oxidation on Pt is a prototypical electrocatalytic reaction which exhibits bistable reaction rates, and thus a bistable current-voltage characteristic.
In technological applications, electrodes consist of billions of catalytically active nanoparticles on a nonreactive support that interact globally when a set current is passed through the electrode.
A setup that enables the measurement of the state of an individual bistable component is an array of Pt-electrodes.
Such measurements reveal key dynamical properties: The bistable units 
activate sequentially upon a slow parameter ramp, they form clusters with most electrodes in the two stable states and at most one electrode on the unstable state of the individual electrode, and they may exhibit collective oscillations of the entire ensemble~\cite{crespo2013cooperative, crespo2014sequential, bozdech2018oscillations}. We recently derived a general necessary condition when a system of bistable units subject to a global constraint can become unstable in a Hopf bifurcation~\cite{salman2020collective} and validated this condition in a simplified CO-electrooxidation model.

\newcommand{\bbS}{\mathbb{S}}
\newcommand{\bbN}{\mathbb{N}}

In this paper, we analyze this CO-electrooxidation model and elucidate its bifurcation properties. 
\hl{The behavior of the coupled units is particularly interesting when the value of the constraint is in the transitional regime between the two outer `active' and  `passive' states of an uncoupled individual unit.
As a globally and identically coupled system of identical units, the bifurcations are constrained by the symmetry properties of the system: The dynamical equations} are $\bbS_N$-symmetric (equivariant), where~$\bbS_N$ denotes the group of permutations of~$N$ symbols that acts by permuting component indices; see Ref.~\onlinecite{Golubitsky2002} for a general introduction to dynamical systems with symmetry. 
The presence of symmetries gives rise to dynamically invariant subspaces, that correspond, for example, to cluster configurations where the state of some components coincide.
Symmetric systems can exhibit bifurcation behavior that is nongeneric for a general dynamical system as their bifurcation behavior is constrained by the symmetry~\cite{Golubitsky2002,Golubitsky1988}.
We will consider symmetry breaking bifurcations where the configuration of all units being synchronized (the one-cluster configuration) loses stability. Indeed, since the state space of individual unit is one-dimensional for the system we consider, collective oscillations cannot occur if all units are synchronized. Thus the emergence of collective oscillations necessarily requires symmetry-breaking bifurcations away from full synchrony.

In our bifurcation analysis we concentrate specifically on the structure imposed by the invariant subspaces due to symmetry. Here the one-, two-, and three cluster subspaces are crucially important as they contain all equilibria of the system. We give a detailed bifurcation analysis of the steady states in these subspaces and highlight bifurcations that give rise to and stabilize limit cycles as they induce collective oscillations observed in experiments. Specifically, the paper is organized as follows. In Section~\ref{sec:Model} we summarize the model equations. In Section~\ref{sec:OneCluster}, we consider the one-dimensional, fully synchronized dynamics where the states of all units take the same value and form one cluster. In Section~\ref{sec:TwoClusters}, we discuss symmetry breaking bifurcations away from full synchrony to two-cluster equilibria {and bifurcations of the two-cluster equilibria}; these can give rise to collective oscillations (that are transversely unstable for most cluster sizes). In Section~\ref{sec:ThreeClusters}, we outline bifurcations of three cluster equilibria and indicate that the transversely unstable collective oscillations within the two-cluster subspace can be stabilized in a transcritical bifurcation. We conclude with some remarks in Section~\ref{sec:Conclusions}.

%--------------------------------------------------------
\section{Symmetric networks of bistable units}
\label{sec:Model}

\subsection{Model equation}

We consider the following simplified model for CO-electrooxidation on Pt-microelectrodes. Its derivation from an established, more detailed model version can be found in the supplement of Ref.~\onlinecite{salman2020collective}. 
The intrinsic state~$x_k$ of unit~$k$ evolves according to
\begin{subequations}\label{eq:ex}
	\begin{align}
	\dt x_k(t) &=\frac{1-(1+ab)x_k(t)}{1+a-x_k(t)}-u(t)[1-x_k(t)]x_k(t)\label{ex1}
	\\
	\y &= \frac{y_\mathrm{tot}}{N} = \frac{1}{N}\sum\limits_{l=1}^N u(t)[1-x_l(t)]x_l(t)\label{ex2}
	\end{align}
\end{subequations}
 {where~$y$ imposes a global constraint} and~$a$ and~$b$ are positive constants; throughout this paper, we fix $a=0.05$, $b=0.01$ as in Ref.~\onlinecite{salman2020collective}. In terms of the CO dynamics,~$x_k$ is the CO-covered fraction of the surface and~$u$ the electrode potential. The macroscopic observable~$y_\mathrm{tot}$ is the total CO oxidation current, and~$y$ denotes the mean current per electrode (or nanoparticle). In our setting,~$y$ is the bifurcation parameter, while~$u$ adjusts such that the system can attain the preset mean value. 
Note that this type of coupling corresponds to a control that fixes the value of a function of the mean state of the ensemble and allows the `natural control parameter'---in our case~$u$---to change. This situation is different from the situation considered in Ref.~\onlinecite{zanette1997dynamics}, where the main bifurcation parameter of the bistable unit is also the bifurcation parameter of the globally coupled ensemble, and thus set to a constant value.

\begin{figure}
	\centering
	\includegraphics[width=\columnwidth]{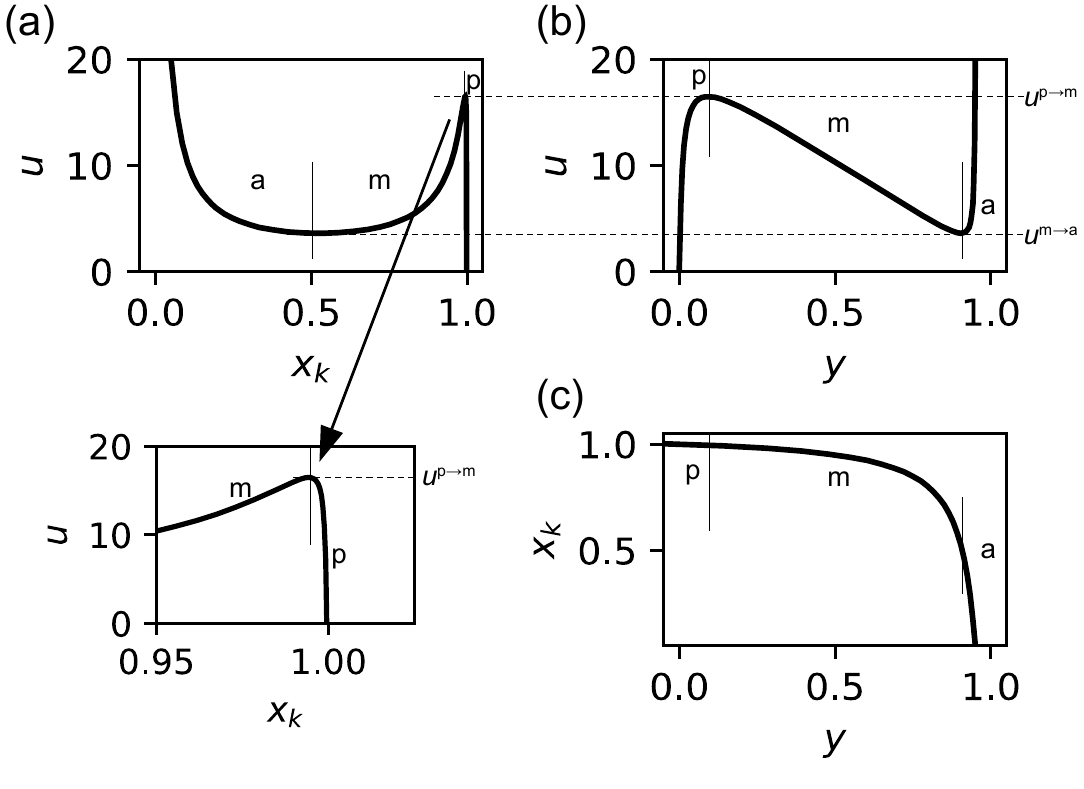}
	\caption{Equilibrium curve of an individual unit, 
	in (a)~the $(u,x_k)$ plane, (b)~the $(u,y)$ plane, (c)~the $(x_k,y)$ plane. The lower left panel shows a magnification of the equilibrium curve in (a).
	For fixed~$u$, the units shows bistable dynamics. The equilibrium branch is subdivided into a passive (``p''), a middle (``m''), and an active (``a'') segment by the saddle-node points at~$u^{\mathrm{p}\to\mathrm{m}}$ and~$u^{\mathrm{m}\to\mathrm{a}}$. These are indicated by labels and vertical lines in all panels. Fixing~$y$ and letting~$u$ adapt allows to select the corresponding segment for an individual unit.
	}
	\label{fig:1d}
\end{figure}

First consider a single uncoupled unit, $N=1$, with state $x_1=x$; the corresponding equilibrium dynamics $\dt x=0$ are shown in Fig.~\ref{fig:1d}.
When~$u$ is treated as a parameter,~$x$ is bistable in~$u$ with equilibrium branch as in Fig.~\ref{fig:1d}a; there are two saddle-node (SN) bifurcations that bound the region of bistability.
For ease of notation, we subdivide the branch of equilibria into three segments specified by the points where these SN bifurcations occur and the branch turns over: The active segment (``a''), where the current is high, the middle (``m'') segment, and the passive segment (``p''), where the current is low (see Fig.~\ref{fig:1d}). We denote the value of~$u$ corresponding to the transitions between the passive and middle segment with $u^{\mathrm{p}\to\mathrm{m}}$, and the transition between the middle and active segment with $u^{\mathrm{m}\to\mathrm{a}}$. The transitions between segments play an explicit role in the bifurcation analysis as we will discuss below.
For the equilibrium values of~$x$, there is a single-valued, but non-monotonic relationship between~$u$ and~$y$, depicted in Fig.~\ref{fig:1d}b.
By contrast the value of~$x$ at equilibrium depends monotonically on~$y$ (Fig.~\ref{fig:1d}c).

The dynamical equations~\eqref{eq:ex} can be rearranged to the explicit form
\begin{equation}
	    \begin{split}
	\label{eq:fullsys}
	\dt x_k(t) =&\frac{1-(1+ab)x_k(t)}{1+a-x_k(t)}
	\\&\quad-\left(\frac{[1-x_k(t)]x_k(t)}{\frac{1}{N}\sum_{l=1}^N [1-x_l(t)]x_l(t)}\right)\y%\label{eq:explicit_x}
	\end{split}
\end{equation}
with the electrode potential
\begin{equation}\label{eq:u}
	u(t) = \frac{\y}{\frac{1}{N}\sum_{l=1}^N [1-x_l(t)]x_l(t)}
\end{equation}
which is determined by the intrinsic states~$x_k$ and the bifurcation parameter~$y$.
If we define the two functions
\begin{subequations}
\begin{align}
P(x) &= \frac{1-(1+ab)x}{1+a-x},\\
Q(x) &= (1-x)x
\end{align}
\end{subequations}
the dynamical equations take the form
	\begin{align}\label{eq:Network}
	\dt x_k(t) &= P(x_k(t)) -\left(\frac{Q(x_k(t))}{\frac{1}{N}\sum_{l=1}^N Q(x_l(t))}\right)\y.
	\end{align}
	\label{eq:fullSys}

Configurations where the states of the units form three clusters are particularly interesting because all equilibria of  Eq.~\eqref{eq:Network} necessarily lie in a three cluster subspace.
To see this, consider the fixed points~$x_k^*$ of Eq.~\eqref{eq:ex} with~$u=u^*$ which satisfy
\begin{equation} \label{eq:poly}
\begin {split}
    0 &= 1-(1+ab)x_k^*-u^*[1-x_k^*]x_k^* (1+a-x_k^*)\\
    &= -u^*(x_k^*)^3 + (2u^*+au^*)(x_k^*)^2 - (1+ab+u^*+au^*)x_k^* + 1. 
\end {split}
\end{equation}
Since this cubic polynomial has at most three distinct roots, the coefficients of any equilibrium of~\eqref{eq:Network} take at most three distinct values and thus form a cluster configuration.

%---------------------------------------------------------
\subsection{Cluster subspaces and full synchrony}

As a globally coupled network of~$N$ identical bistable units, the system equation~\eqref{eq:Network} are symmetric (equivariant) with respect to permutations of units. In other words, if the group~$\bbS_N$ of permutations of~$N$ symbols acts on the bistable units by permuting indices, the equations of motion remain unchanged. This implies that cluster configurations, where the state of all units in each cluster is identical, are also invariant under the dynamics.

Suppose that the units form~$M$ clusters such that the $k$th~cluster contains~$N_k$ units with identical state~$\xi_k$. Evidently, we have to have $N=\sum_{j=1}^MN_j$.
Define the relative cluster size $n_k = \frac{N_k}{N}$ and write $C_{(n_1,\dotsc,n_M)}$
for the cluster subspace where the first~$N_1$ oscillators form cluster~$1$, the next $N_2$ oscillators cluster~$2$, and so on.
The effective dynamics on~$C_{(n_1,\dotsc,n_M)}$---and by symmetry any other cluster configuration with these cluster sizes---are $M$-dimensional and the state~$\xi_k$ of cluster $k\in\{1,\dotsc,M\}$ evolves according to
	\begin{align}
	\dt \xi_k(t) &= P(\xi_k) -\left(\frac{Q(\xi_k(t))}{\sum_{j=1}^M n_j Q(\xi_j(t))}\right)\y.
	\end{align}
While~$n_k$ only takes finitely many values for finite networks, in the limit of infinitely many units, $N\to\infty$, we can see~$n_k$ as a continuous parameter.

%--------------------------------------------------------
\section{Fully synchronized dynamics}
\label{sec:OneCluster}

The simplest cluster configuration~$C_{(1)}$ is a single cluster, that is, all units are synchronized with $x_k=\xi_1=\xi$.
With $P,Q$ as above, the synchronized dynamics of~\eqref{eq:Network} are given by
	\begin{align}\label{eq:1d_dyn}
	\dt \xi(t) &=\frac{1-(1+ab)\xi(t)}{1+a-\xi(t)}
	-\y.
	\end{align}
Eq.~\eqref{eq:1d_dyn} has only one single equilibrium point
	\begin{align}
	\xi^* &= \frac{1-\y-a\y}{1+ab-\y}\label{eq:1d}
	\end{align}
for any~$\y\in(0,1)$, which is identical to the curve of an individual element depicted in Fig.~\ref{fig:1d}.
Restricted to the synchronized subspace, the system is monostable for all parameter values, i.e., for each~$y$ there is exactly one single equilibrium.
(If~$u$ is given instead of~$y$, however, there can be three equilibria.) Linearizing Eq.~\eqref{eq:1d_dyn} at the equilibrium~$\xi^*$ yields
\begin{equation}
\dt \xi(t) = \xi^* - \frac{(1-y+ab)^2}{a(ab+b+1)}\xi(t)+\dots
\end{equation}
and thus~$\xi^*$ is always stable within~$C_{(1)}$.

The stability of~$\xi^*$ as an equilibrium of the full system~\eqref{eq:Network} depends on the~$N-1$ eigenvalues transverse to~$C_{(1)}$. Due to the symmetry, all transversal eigenvalues are the same; 
they are all negative on the ``a'' and ``p" segments, and positive on the ``m" segment. 
This implies that the equilibrium is stable \hl{on the ``a'' and ``p'' segments
 and unstable on the ``m'' segment. The change of stability of the equilibrium~$\xi^*$ happens in} $\bbS_N$-equivariant transcritical bifurcations at {$u^{\mathrm{p}\to\mathrm{m}}$ and~$u^{\mathrm{m}\to\mathrm{a}}$}
where all eigenvalues transverse to~$C_{(1)}$ pass through zero simultaneously. 
\hl{Note that at these transcritical bifurcations an uncoupled individual bistable unit undergoes a SN bifurcation} 
if~$u$ is considered as a bifurcation parameter (cf.~Eq.~\eqref{ex1}). That is, the bifurcation happens where all synchronized units jointly transition from the passive to the middle segment of the branch or from the middle to the active segment of the branch.

%--------------------------------------------------------
\section{Two-cluster dynamics and bifurcations}
\label{sec:TwoClusters}

Now, we consider two-cluster configurations~$C_{(n_1, n_2)}$ with relative sizes~$n_1$ and~$n_2$, respectively. The two-dimensional dynamics on $C_{(n_1, n_2)}$ are given by
\begin{subequations}\label{eq:TwoClusterDyn}
	\begin{align}
	\dt \xi_1(t) &= P(\xi_1(t)) -\left(\frac{Q(\xi_1(t))}{ n_1 Q(\xi_1(t)) +  n_2 Q(\xi_2(t))}\right)\y \label{eq:2d_ode1}\\
	\dt \xi_2(t) &= P(\xi_2(t)) -\left(\frac{Q(\xi_2(t))}{ n_1 Q(\xi_1(t)) +  n_2 Q(\xi_2(t))}\right)\y.\label{eq:2d_ode2}
	\end{align}
\end{subequations}
The intersection of all possible two-cluster subspaces contain the fully synchronized subspace $\{\xi_1=\xi_2\}$. 
Note that for~$n_1$ and~$n_2$ fixed, the system has a parameter symmetry 
\[\left(n_1, n_2; \xi_1, \xi_2\right)\mapsto \left(n_2, n_1; \xi_2, \xi_1\right),\] 
which corresponds to exchanging the clusters. 
Since $n_1+n_2=1$ the parameter~$n_1$ fully determines the system and the parameter symmetry can also be written as \begin{equation}\label{eq:ParamSym}
    \left(n_1; \xi_1, \xi_2\right)\mapsto \left(1-n_1; \xi_2, \xi_1\right).
\end{equation}
Thus, without loss of generality, we may assume $n_1\geq n_2$. 
If the clusters are of equal size, $n_1 = n_2$, the parameter symmetry yields a system symmetry $(\xi_1, \xi_2) \mapsto (\xi_2, \xi_1)$ on the cluster subspace.

\begin{figure}
	\centering
	\includegraphics[width=\columnwidth]{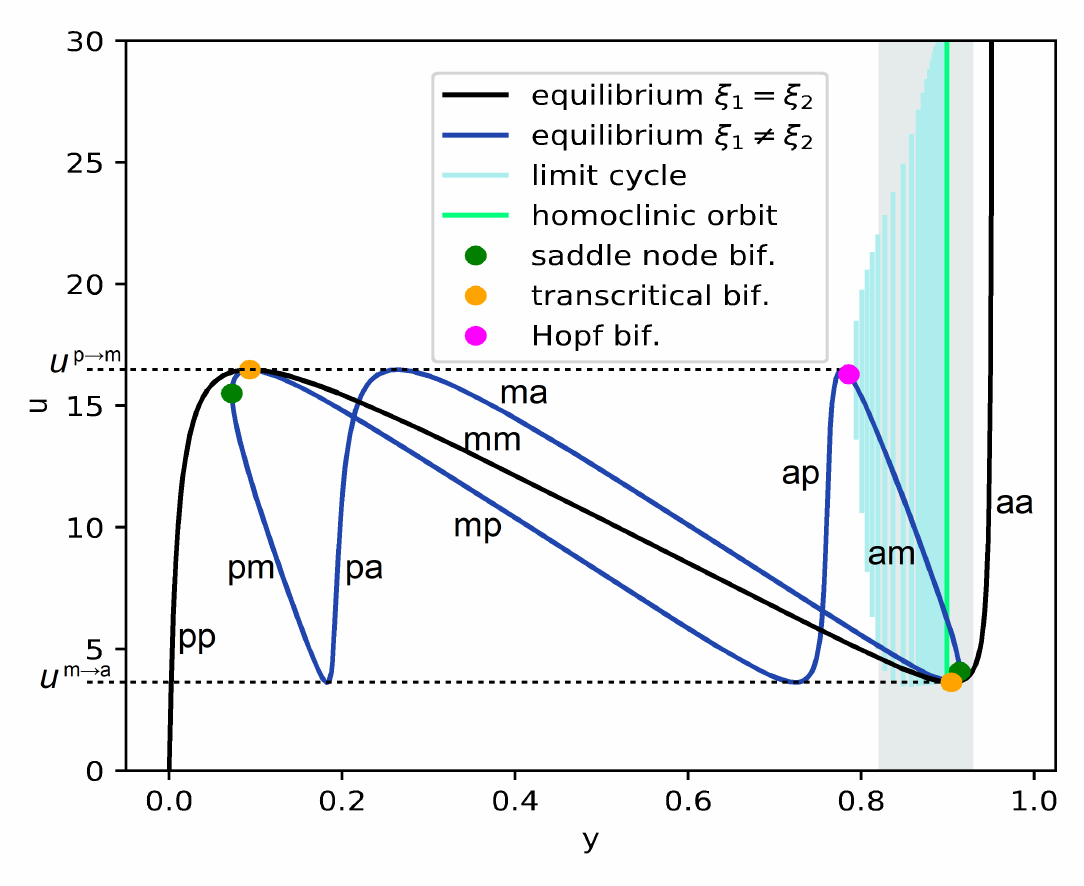}
	\caption{Symmetry breaking bifurcations and transitions to periodic dynamics occur on the two-cluster subspace $C_{ (n_1, n_2) }$ for $n_1=0.8$, $n_2=0.2$, for $a=0.05$, $b=0.01$.
    The black line indicates the branch of fully symmetric equilibria as in Fig.~\ref{fig:1d}. This branch interacts with the two-cluster equilibria (blue line) in transcritical bifurcations (yellow dot). The labels ``a'', ``m'', ``p'' indicate the active, middle, and passive segments of the branch as detailed in the main text with the first letter representing~$\xi_1$, the second~$\xi_2$. A Hopf bifurcation (green dot) gives rise to oscillatory dynamics. The shaded parameter range is considered in detail in Fig.~\ref{fig:2d_phasePortrait}.
	}
	\label{fig:2d_eq}
\end{figure}

The branch of fully symmetric equilibria in~$C_{(1)}$ interacts with branches of two-cluster solutions in~$C_{(n_1, n_2)}$ in transcritical bifurcations. {Note that we say that two (distinct) branches of equilibria or limit cycles \textit{interact} if they are involved in the same bifurcation; for example, two branches of equilibria interact in a simple transcritical bifurcation.} All branches and their bifurcations described throughout the paper were calculated numerically using AUTO-07P~\cite{doedel2007auto}. 
In the following, we will 
{first} focus on the specific relative cluster sizes $n_1=0.8$, $n_2=0.2$ to illustrate the bifurcation behavior.
Fig.~\ref{fig:2d_eq} shows the equilibria branches in the $(u,y)$-plane as the bifurcation parameter~$y$ is varied.
The black line shows the fully symmetric equilibrium $\xi^= = (\xi^*,\xi^*)$---with $\xi^*$ as in Section~\ref{sec:OneCluster} 
---within $C_{(1)}\subset C_{(n_1, n_2)}$; this branch is the same as the one shown in Fig.~\ref{fig:1d}b.

The locations where the fully symmetric equilibrium solution undergoes transcritical bifurcations are indicated by yellow dots. At these bifurcation points, the symmetric branch interacts with the branch of two-cluster equilibria (dark blue line).
\hl{It is useful to subdivide the branch of two-cluster equilibria into segments: We write ``aa'' if both clusters are on the active segment,
``am'' if the first cluster is on the active and the second one on the middle segment, 
etc. This leads to the labelling shown in Fig.~\ref{fig:2d_eq} and the boundaries of the segments are given by the local maxima and minima corresponding to $u\in \{u^{ \mathrm{p}\to\mathrm{m}},u^{\mathrm{m}\to\mathrm{a} }\}$. 
This labelling yields some intuition of what happens at the transcritical bifurcations: At the first transcritical bifurcations the two-cluster equilibrium transitions from ``pm'' to ``mp'' and the clusters switch roles as the majority cluster ($n_1=0.8$) switches from ``p'' to ``m'' and the minority cluster ($n_2=0.2$) from ``m'' to ``p''. Similarly, at the second transcritical bifurcation there is a transition from ``ma'' to ``am''.}
Following the two-cluster equilibrium branch further, it folds over twice in SN bifurcations (green dots) and undergoes a Hopf bifurcation at $y\approx 0.8$ (purple dot).
This gives rise to a branch of limit cycles (light blue) that is stable within~$C_{(n_1, n_2)}$ and ends in a homoclinic bifurcation close to the transcritical bifurcation at $y\approx0.9$.

\begin{figure*}
	\centering
	\includegraphics[width=16cm]{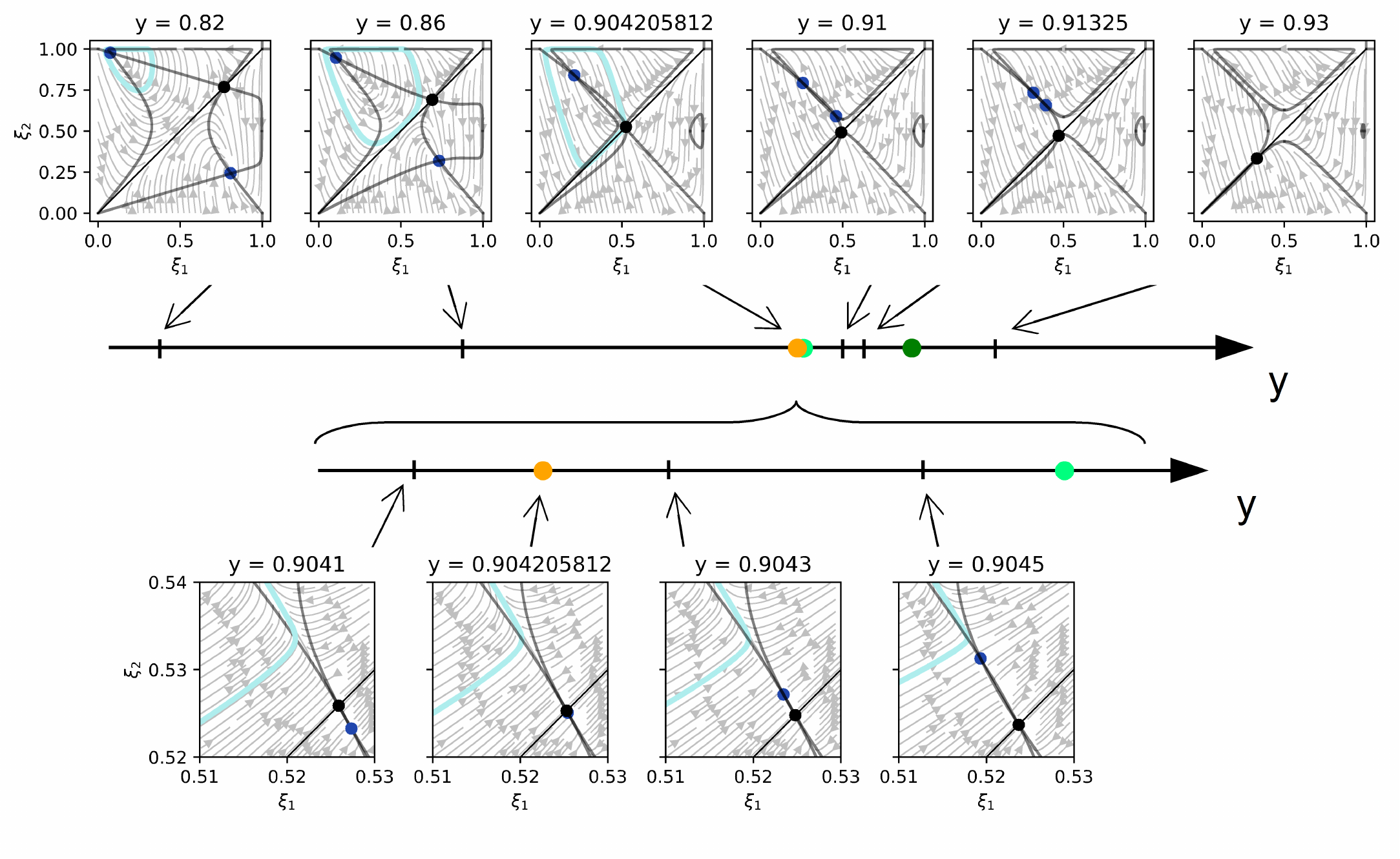}
	\caption{Phase portraits elucidate {the dynamics in $C_{(n_1,n_2)}$ due to the successive transcritical bifurcation (yellow dot), homoclinic bifurcation (light green dot), and saddle node bifurcation (green dot)} for $n_1=0.8$, $n_2=0.2$, for $a=0.05$, $b=0.01$. 
    Gray solid lines indicate the nullclines; solutions and bifurcation points are colored as in Fig.~\ref{fig:2d_eq}. As the parameter~$y$ is increased, the limit cycle (light blue circle) that arises from the Hopf bifurcation of~$\xi^+$ (upper blue dot in the first panel) grows quickly. At the same time, the two-cluster equilibrium~$\xi^-$ (lower blue dot in the first panel) crosses through the invariant subspace~$C_{(1)}$---the diagonal in the $(\xi_1,\xi_2)$ plane---in the transcritical bifurcation before it interacts with the limit cycle in a homoclinic bifurcation. Finally, the equilibria~$\xi^+$ and~$\xi^-$ merge in a saddle node bifurcation.
	}
	\label{fig:2d_phasePortrait}
\end{figure*}

There is actually a series of bifurcations close to the second transcritical bifurcation as the parameter~$y$ is varied as shown in Fig.~\ref{fig:2d_phasePortrait}. When the Hopf bifurcation occurs at $y\approx 0.8$, there are three equilibria in $C_{(n_1, n_2)}$, the symmetric equilibrium $\xi^=$ in $C_{(1)}\subset C_{(n_1, n_2)}$ as well as nonsymmetric equilibria $\xi^+=(\xi^+_1,\xi^+_2)$ with $\xi^+_1<\xi^+_2$ and $\xi^-=(\xi^-_1,\xi^-_2)$ with $\xi^-_2<\xi^-_1$; the former gives rise to the stable limit cycle. At the transcritical bifurcation~$\xi^-$ interacts with ~$\xi^=$ and satisfies $\xi^-_1<\xi^-_2$ after the bifurcation point. This equilibrium then collides with the limit cycle in a homoclinic bifurcation at $y\approx0.91$. Finally, the two equilibria $\xi^-$ and $\xi^+$ meet in the fold where the nonsymmetric branch in Fig.~\ref{fig:2d_eq} folds over itself. Note that the transcritical bifurcation must happen before the homoclinic bifurcation to allow for this bifurcation scenario: {It cannot occur on the ``ma'' segment before the transcritical bifurcation but has to occur on the ``am'' segment (but before~$\xi^-$ and~$\xi^+$ merge in the SN bifurcation).}

Considering transverse stability beyond~$C_{(n_1, n_2)}$ 
we realize that the branch of two-cluster equilibria undergoes transcritical bifurcations whenever $u\in \{u^{\mathrm{p}\to\mathrm{m}},u^{\mathrm{m}\to\mathrm{a}}\}$. Hence, there are six in total, the two indicated by yellow dots in Fig.~\ref{fig:2d_eq}, where all bifurcating branches lie within~$C_{(n_1, n_2)}$, and four additional ones where the two-cluster equilibria interact with three-cluster equilibria; we will discuss these in more detail in the next section.

\begin{figure}
	\centering
	\includegraphics[width=\columnwidth]{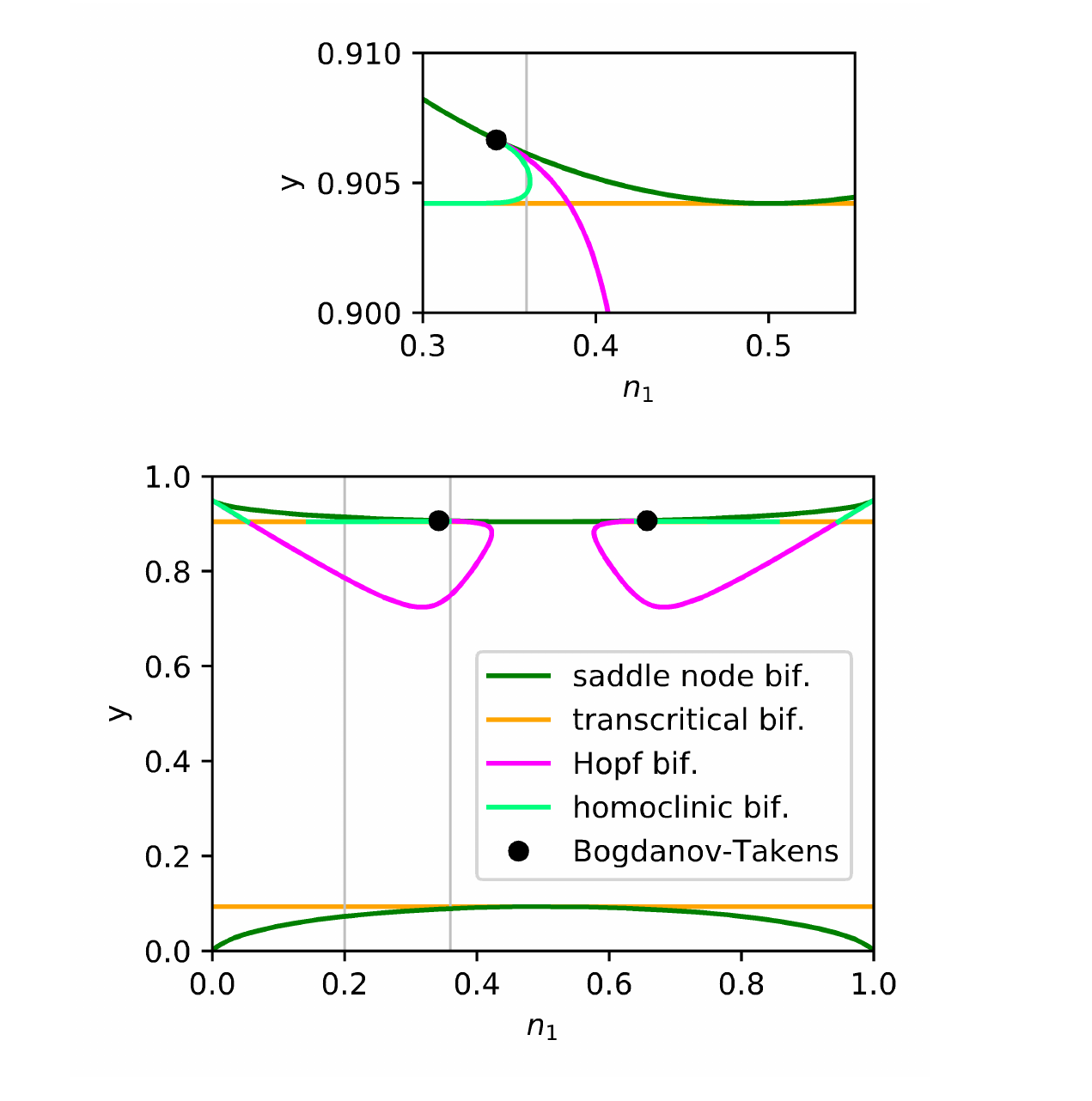}
	\caption{A codimension-2 Bogdanov--Takens bifurcation organizes the bifurcation behavior in~$C_{ (n_1,n_2) }$ as both~$y$ and the relative cluster size~$n_1$ is varied; the remaining parameters are $n_2=1-n_1$, $a=0.05$, and $b=0.01$. The symmetry of the bifurcations is due to the parameter symmetry~\eqref{eq:ParamSym}.
	The upper panel shows a magnification close to the codinmension-2 Bogdanov--Takens point where saddle-node, Hopf, and homoclinic bifurcation branches meet.
	For $n_1=n_2=\frac{1}{2}$, the saddle node and transcritical bifurcations merge in a pitchfork symmetry breaking bifurcation.
	}
	\label{fig:2d_map}
\end{figure}

To elucidate the dynamics and bifurcations for other relative cluster sizes, we continued the bifurcations in two parameters, $y$ and~$n_1$. The resulting two-parameter bifurcation lines are shown in Fig.~\ref{fig:2d_map}. Note that the symmetry of the bifurcation diagram reflects the parameter symmetry~\eqref{eq:ParamSym}. The transcritical bifurcation lines are horizontal {since the bifurcation condition does not depend on the relative cluster sizes:
Consider equilibria on~$C_{(n_1,n_2)}$, that is, solutions of Eq.~\eqref{eq:poly} with two distinct real roots~$\xi^*_1\neq\xi^*_2$ (a nonsymmetric equilibrium on~$C_{(n_1,n_2)}$). Now suppose that these roots coincide at a bifurcation for $u=u^*$, defined in~\eqref{eq:u}, which means that~\eqref{eq:poly} has a triple root~$\xi^*_1=\xi^*_2=\xi^*$ (an equilibrium on $C_{(1)}\subset C_{(n_1,n_2)}$). Then, the associated value of~$\y$ is
\begin{align}\label{eq:transcrit_2c}
\y &= n_1 u^*Q(\xi_1) + n_2 u^*Q(\xi_2) = u^*Q(\xi^*),
\end{align}
independent of~$n_1$.
}
Furthermore, for any given relative cluster sizes~$(n_1, n_2)$, there is a branch of equilibria in~$C_{(n_1, n_2)}$.
These branches arise in saddle-node bifurcations and interact with the one-cluster equilibrium in the transcritical bifurcation. The more asymmetric the clusters are, the larger is the distance of the saddle-node bifurcation to the transcritical bifurcation in the bifurcation parameter~$y$. 
 The Hopf and homoclinic bifurcations of the two-cluster states described above extend to a large range of cluster sizes from the most asymmetric distribution~$n_1=1$ (or, equivalently,~$n_1=0$) to~$n_1\approx 0.4$  (respectively $n_1\approx 0.6$). They meet in a codimension-two Bogdanov--Takens bifucation point at~$(y,n_1)\approx(0.907,0.34)$ ($(y,n_1)\approx(0.907,0.66)$, respectively). For increasing (respectively decreasing) cluster size~$n_1$, the Hopf bifurcation curve approaches the saddle node bifurcation curve again. In this region of parameter space, the limit cycle solution grows extremely quickly as in a Canard phenomenon~\cite{Krupa2001} before it is destroyed in a branch of homoclinic bifurcations.

In the full system~\eqref{eq:Network} all these bifurcations happen simultaneously in a large number of invariant subspaces. 
On the one hand, for fixed~$(n_1,n_2)$ the bifurcations happen simultaneously in the symmetric copies of~$C_{(n_1, n_2)}$.
On the other hand, recall that~$(n_1,n_2)$ parameterize invariant subspaces of different cluster sizes. {The transcritical bifurcations for all the different~$(n_1,n_2)$ happen at the same value of~$y$ given by~\eqref{eq:transcrit_2c}. Thus, the} symmetric equilibrium undergoes a transcritical bifurcation simultaneously in all these invariant subspaces as one would expect in our symmetric setting. In the case of symmetric clusters, $n_1=n_2=0.5$, these degenerate to a symmetric pitchfork bifurcation, in line with the reflection symmetry $(\xi_1,\xi_2)\mapsto(\xi_2,\xi_1)$ on~$C_{(n_1, n_2)}$; cf.~Fig.~\ref{fig:2d_map}.

%-------------------------------------------------------------
\section{Three-cluster dynamics and bifurcations}
\label{sec:ThreeClusters}

Now consider the three-cluster subspace $C_{(n_1, n_2, n_3)}$ with three clusters of relative sizes $n_1+n_2+n_3=1$ whose states $\xi_1,\xi_2,\xi_3$ evolve according to
	\begin{subequations}\label{eq:ThreeClusterDyn}
		\begin{align}
		\dot \xi_1 &= P(\xi_1) -\left(\frac{Q(\xi_1)}{ n_1 Q(\xi_1) +  n_2 Q(\xi_2) +  n_3 Q(\xi_3)}\right)\y\label{eq:3d_dyn1}\\
		\dot \xi_2 &= P(\xi_2) -\left(\frac{Q(\xi_2)}{ n_1 Q(\xi_1) +  n_2 Q(\xi_2) +  n_3 Q(\xi_3)}\right)\y\\
		\dot \xi_3 &= P(\xi_3) -\left(\frac{Q(\xi_3)}{n_1 Q(\xi_1) +  n_2 Q(\xi_2) +  n_3 Q(\xi_3)}\right)\y.\label{eq:3d_dyn4}
		\end{align}
\end{subequations}%}
The invariant three-cluster space~$C_{(n_1,n_2,n_3)}$ contains the invariant two-cluster spaces $C_{(n_1+n_2, n_3)}$ as $\{\xi_1 = \xi_2\}$, $C_{(n_1, n_2+n_3)}$ as $\{\xi_2 = \xi_3\}$, and $C_{(n_1+n_3,n_2)}$ as $\{\xi_3 = \xi_1\}$,
on which the dynamics are given by~\eqref{eq:TwoClusterDyn}. 
As above, any permutation of the three clusters yields a parameter symmetry. 
If $n_1 = n_2$, $n_2=n_3$, or $n_3=n_1$, then~\eqref{eq:ThreeClusterDyn} has a symmetry 
with respect {to the transpositions that swap the corresponding clusters.}
Thus, without loss of generality we may assume $n_1\geq n_2\geq n_3$. 
If $n_1=n_2=n_3=\frac{1}{3}$ (i.e., all clusters are of equal size) then the equations~\eqref{eq:ThreeClusterDyn} are equivariant with respect to the full permutation of three elements.

%------------------------------------
\subsection{Equivariant bifurcation of equilibria}

\begin{figure}
	\centering
	\includegraphics[width=\columnwidth]{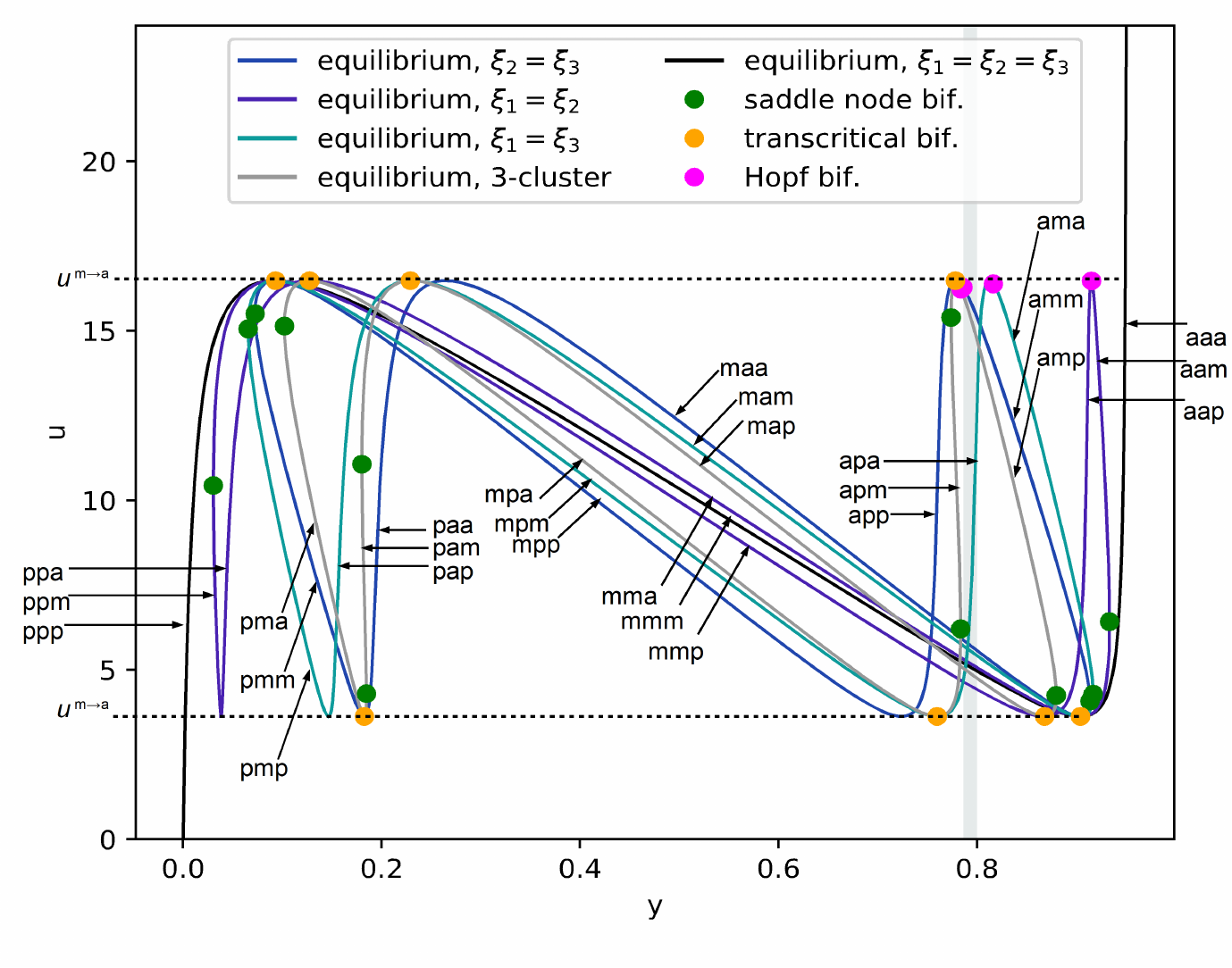}
	\caption{
	    Three-cluster subspaces $C_{(n_1, n_2, n_3)}$ contain all possible equilibrium curves; here, branches of solutions in~$C_{(n_1, n_2, n_3)}$ for relative cluster sizes $n_1=0.8$, $n_2=0.16$, $n_3=0.04$ are shown as~$y$ is varied and $a=0.05$, $b=0.01$ are fixed. 
		Like in Fig.~\ref{fig:1d} and explained in the main text, the labeling with letters ``a'', ``m'', ``p'' indicate the active, middle, and passive segments of the equilibrium branch in cluster~1, 2, and 3, respectively.
	    The black line shows the fully symmetric equilibrium (as in Figs.~\ref{fig:1d} and~\ref{fig:2d_eq}), blue lines of different color the three different two-cluster equilibrium branches (the dark blue with $\xi_2=\xi_3$ is the same one shown in Fig.~\ref{fig:2d_eq}), and the gray line corresponds to a branch of asymmetric equilibria. Two- and three-cluster equilibria interact in symmetric transcritical bifurcations (yellow dots) and other bifurcation points are indicated as above.
        {The shaded parameter range is considered in detail in Fig.~\ref{fig:3d_pp_transcrit}.}	
}
	\label{fig:3d_equlibrium}
\end{figure}

Since any equilibrium is constrained to a three-cluster configuration due to condition~\eqref{eq:poly}, any branch of steady states is contained in a cluster subspace~$C_{(n_1, n_2, n_3)}$. 
Fig.~\ref{fig:3d_equlibrium} shows the equilibrium branches in~$C_{(n_1, n_2, n_3)}$ for relative cluster sizes $n_1=0.8$, $n_2=0.16$, and $n_3= 0.04$.
We have $C_{(1)}\subset C_{(n_1, n_2, n_3)}$, so the branch of symmetric equilibria in Fig.~\ref{fig:1d} appears (black line).
Note that $n_2+n_3=0.2$ and thus $C_{(0.8, 0.2)}\subset C_{(0.8, 0.16, 0.04)}$ for $\{\xi_2=\xi_3\}$ and the equilibrium branches shown in Fig.~\ref{fig:2d_eq} reappear in Fig.~\ref{fig:3d_equlibrium} (blue line).
In addition, $C_{(0.8, 0.16, 0.04)}$~also contains the two-cluster subspaces $\{\xi_1=\xi_3\}$ and $\{\xi_1=\xi_2\}$; these correspond to~$C_{(0.84, 0.16)}$ and~$C_{(0.96, 0.04)}$ respectively. The steady state bifurcations can be read in Fig.~\ref{fig:2d_map} and the corresponding branches of equilibria are depicted in Fig.~\ref{fig:3d_equlibrium} as purple and teal lines, respectively.
Finally, there is one branch of three-cluster equilibria. Taken together, for $u \in [u^{\mathrm{p}\to\mathrm{m}},u^{\mathrm{m}\to\mathrm{a}}]$  there are $3^3=27$ equilibria because each~$x_j$ can take one of three values while $\dot x_j=0$, $\dot u=0$). 

These equilibria now (ex)change their stability properties as they bifurcate.
In addition to the transcritical bifurcations of the fully synchronized equilibrium, the two-cluster equilibria also undergo transcritical bifurcations in which three-cluster equilibria are involved.
More specifically, each of the three two-cluster subspaces contained in~$C_{(0.8, 0.16, 0.04)}$, that is, $\xi_1=\xi_2$, $\xi_2=\xi_3$, $\xi_1=\xi_3$, contains a transcritical bifurcation for each of the two double root of the equilibrium equation~\eqref{eq:poly}.
This yields six transcritical bifurcations in addition to transcritical bifurcations of the fully synchronized equilibrium that already appeared in Fig.~\ref{fig:2d_eq}.
These bifurcations link to the physical interpretation as in the previous section:
At each transcritical bifurcation, the value of~$u$ is identical with one of the values at which the individual unit undergoes a saddle node bifurcation when~$u$ is considered as a bifurcation parameter.
Indeed, 
it can be shown that when the transcritical bifurcation involves an equilibrium with the larger of the two $u$-values, i.e., $u = u^{\mathrm{p}\to\mathrm{m}}$ the components of the three-cluster state which are on the middle and passive branch segment switch the segment
while the component on the active segment remains unchanged.
For example, if $\xi^{\mathrm{x}}_j$, $j\in\{1,2,3\}$ denotes the $j$th component of a three cluster that is on the segment $\mathrm{x}\in\{\mathrm{p},\mathrm{m},\mathrm{a}\}$, then an equilibrium $\xi^*=(\xi^\mathrm{a}_1,\xi^\mathrm{m}_2,\xi^\mathrm{p}_3)$ on the ``amp'' segment transitions to $\xi^*=(\xi^\mathrm{a}_1,\xi^\mathrm{p}_2,\xi^\mathrm{m}_3)$ on the ``apm'' segment. Correspondingly, at the lower value of~$u$ the components with values on the middle and active segment switch segment while the one on the passive segment remains there, such as $\xi^*=(\xi^\mathrm{a}_1,\xi^\mathrm{m}_2,\xi^\mathrm{p}_3)$ transitions to $\xi^*=(\xi^\mathrm{m}_1,\xi^\mathrm{a}_2,\xi^\mathrm{p}_3)$.

\begin{figure}
	\centering
	\includegraphics[width=\columnwidth]{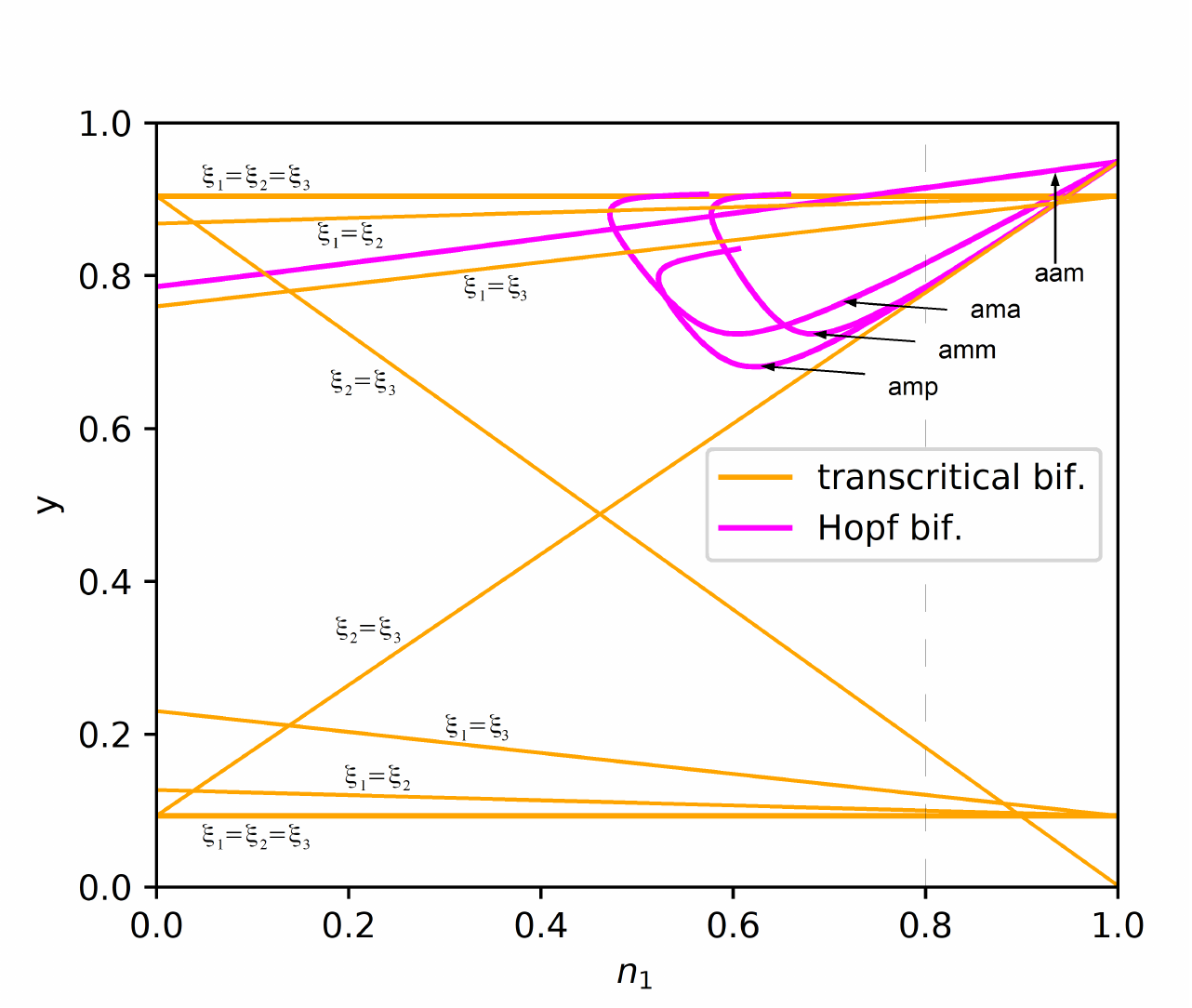}
	\caption{
	    Numerical continuation shows two-parameter bifurcation as both the bifurcation parameter~$y$ and the relative cluster sizes are varied. Here transcritical bifurcation branches (yellow lines; cf.~Eq.~\eqref{eq:transcrit}) as well as Hopf bifurcation branches (purple lines) occurring in three-cluster subspaces are shown for the relative cluster sizes $n_2=0.8\cdot(1-n_1)$, $n_3=0.2\cdot(1-n_1)$ and fixed parameter $a=0.05$, $b=0.01$. The letters ``a'', ``m'', ``p'' indicate how the clusters distribute on the three segments (see Fig.~\ref{fig:1d}) of the equilibria that undergo the Hopf bifurcation.}
	\label{fig:3d_map}
\end{figure}

{We can evaluate the values at which the transcritical bifurcations occur explicitly as in Sect.~\ref{sec:TwoClusters}; in general, they depend on the cluster sizes~$n_1$, $n_2$,~$n_3$.
Consider equilibria on~$C_{(n_1,n_2,n_3)}$, that is, solutions of Eq.~\eqref{eq:poly} with three distinct real roots~$\xi^*_1\neq\xi^*_2\neq\xi^*_3\neq\xi^*_1$ (a nonsymmetric equilibrium on~$C_{(n_1,n_2,n_3)}$). Now suppose two of these roots coincide at a bifurcation for $u=u^*$, defined in~\eqref{eq:u} so that Eq.~\eqref{eq:poly} has one real double root~$\xi^\mathrm{d}$ and one real single root $\xi^\mathrm{s}$. Without loss of generality we assume that $\xi^*_1=\xi^\mathrm{s}$, $\xi^*_2=\xi^*_3=\xi^\mathrm{d}$ on $C_{(n_1, n_2+n_3)}\subset C_{(n_1, n_2,n_3)}$. Then, the associated value of~$\y$ is
	\begin{equation}\label{eq:transcrit}
	\begin{split}
	\y &= n_1 u^*Q(\xi^*_1) + n_2 u^*Q(\xi^*_2)+ n_3 u^*Q(\xi^*_3)\\
	&= 
	u^*[Q(\xi^\mathrm{s})-Q(\xi^\mathrm{d})] n_1 + u^*Q(\xi^\mathrm{d}),
	\end{split}
	\end{equation}
which is linear in~$n_1$.}
The cluster at $\xi^\mathrm{s}$ is what distinguishes this from the previous lower-dimensional case, where this ``bystander'' was not present.
By symmetry, the condition for the transcritical bifurcation in the other two-cluster subspaces depends linearly on $n_2$, $n_3$, respectively.
Thus the transcritical bifurcations are given by lines as the cluster sizes are varied, shown in Fig.~\ref{fig:3d_map}.
In this figure, we see that eight transcritical bifurcations exist for every~$n_1$. The value $n_1=0.8$ corresponds to Fig.~\ref{fig:3d_equlibrium}. We see the two transcritical bifurcations in $C_{(n_1, n_2+n_3)}$ following Eq.~\eqref{eq:transcrit}. Furthermore, $n_2$ and~$n_3$ are chosen to be linearly dependent on~$n_1$ (see caption Fig.~\ref{fig:3d_map}), so the corresponding bifurcations lie on straight lines, too.

For the full system with~$N$ identical units this means that for each $\ell\in\{2,\dots,N\}$ (i.e., the combined size of cluster 2 and 3, which fall together at the bifurcation) there are $\binom{N}{\ell}\sum_{k=1}^{\ell-1}\frac{1}{2}\binom{\ell}{k}$ transcritical bifurcations at $\y = u^*[Q(\xi^\mathrm{s})-Q(\xi^\mathrm{d})]\,(N-\ell)/N+u^*Q(\xi^\mathrm{d})$.
These bifurcation points lie in~$\binom{N}{\ell}$ different subspaces, depending on which~$x_k$ are at~$\xi^\mathrm{s}$ and~$\xi^\mathrm{d}$.
If you consider bifurcations as identical when they can be switched between by index permutation, then there is only one such bifurcation for each combination of $\ell\in\{2,\dots,N\}$ (i.e., combined size of cluster 2 and 3) and $k\in\{1,\dots,\ell-1\}$ (i.e., size of cluster 2).
Furthermore, many of these transcritical bifurcations lie on the same points in parameter space and in phase space such that in total there are actually just $N-1$ multi-branch bifurcations-points at every~$u^*$, i.e. one for each possible size of cluster~1 (the cluster at $\xi^\mathrm{s}$), i.e., one for each $\ell\in\{2,\dots,N\}$. 
The number of branches is $2 + 2\sum_{k=1}^{\ell-1}\frac{1}{2}\binom{\ell}{k}$.
These bifurcations are invariant under intra-cluster index-permutations and are very similar to the equivariant bifurcations we saw in lower dimensions. Here, however, they can occur at various different values of~$y$. In fact, for $N \to \infty$ they occur at each of the two $u^*$-values in the entire $y$-intervals from the transcritical bifurcations almost to the respective fully symmetric state.  

As seen above for the two-cluster subspaces, also three-cluster equilibria might undergo a Hopf bifurcation. In Fig.~\ref{fig:3d_map} the locations of the Hopf-bifurcations of the three-cluster and the different two-cluster equilibria are depicted in the $(y,n_1)$-plane. The labels of the different curves indicate the segments of the three components of the equilibria from which the limit cycles bifurcated. All of them contain at least one cluster on the middle segment. As shown in Ref.~\onlinecite{salman2020collective}, in the full system, any steady state with more than one element on the middle branch is unstable. Hence, even if the Hopf-bifurcations are supercritical and the limit cycle branch off equilibria that are stable in the cluster subspaces, they are unstable in the full system. The limit cycles which emerge in a Hopf bifurcation will therefore not be observable in systems containing a large number of individual elements. However, as we will demonstrate in the next subsection, the three-cluster limit cycles might interact with a two-cluster limit cycle in a transcritical bifurcation, thereby stabilizing the limit cycle in the two-cluster subspace.

\subsection{Equivariant bifurcation of limit cycles}

\begin{figure}
	\centering
	\includegraphics[width=\linewidth]{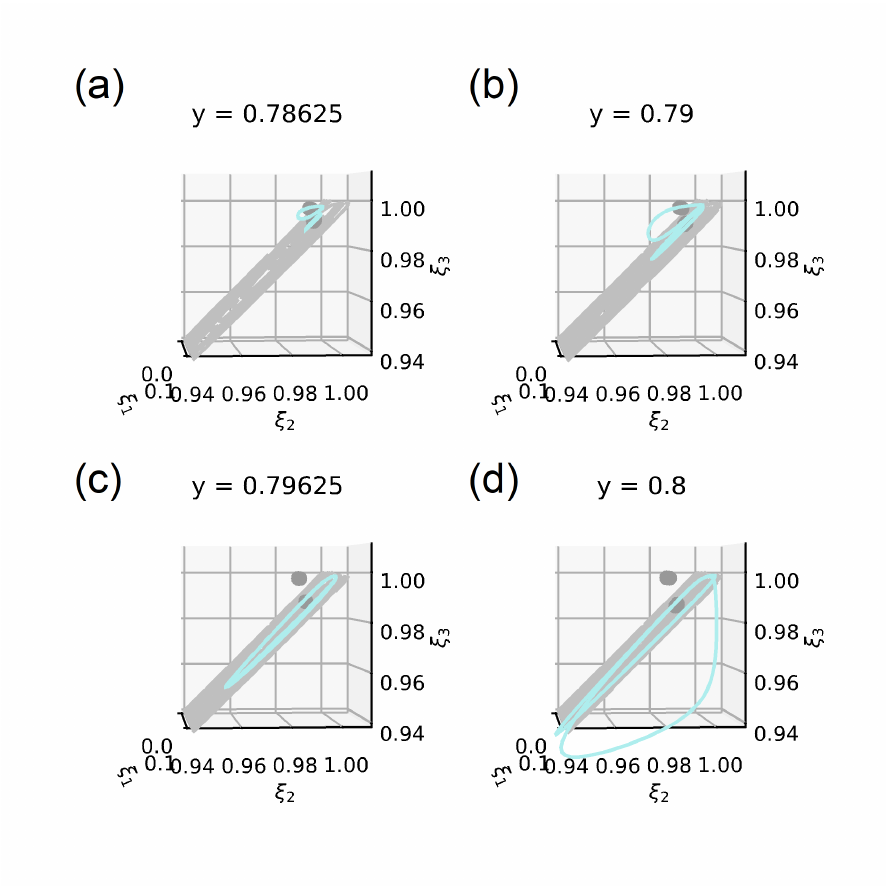}
	\caption{Variation of the parameter~$y$ indicates a transcritical bifurcation of limit cycles as illustrated in phase portraits; the other  parameters are $n_1=0.8$, $n_2=0.16$, $n_3=0.04$, $a=0.05$, $b=0.01$. 
	(a): Two limit cycles have just branched from a Hopf bifurcation at slighly smaller $y$. One of these two limit cycles lies on a plane, which is the two-cluster subspace $C_{(n_1, n_2+n_3)}\subset C_{(n_1, n_2, n_3)}$. The other lies above the plane in the three cluster subspace $C_{(n_1, n_2, n_3)}$. (b): The two limit cycles grow as $y$ is further increased. (c): The two limit cycles interact in a tanscritical bifurcation of limit cycles. (d): The three-cluster limit cycle is now below the two-cluster plane.
	}
	\label{fig:3d_pp_transcrit}
\end{figure}

Consider the phase portraits in the three-cluster subspace~$C_{(0.8,0.16,0.04)}$ depicted in Fig.~\ref{fig:3d_pp_transcrit}. At $y=0.78625$, there are two two fixed points and two limit cycles which were both born at slightly smaller values of~$y$ in supercritical Hopf bifurcations. One limit cycle and one fixed point are three-cluster limit sets, while the other two limit sets lie in the invariant~$C_{(n_1, n_2+n_3)}$ subspace, i.e., $\xi_2=\xi_3$.
The two-cluster limit cycle is stable within the two-cluster subspace but unstable in the direction perpendicular to it, while the three-cluster limit cycle is stable within the invariant three-cluster subspace. 
As~$y$ is increased to $y=0.79$, we see that the limit cycles have increased in size and have moved closer together.
At $y\approx0.79625$ there is a transcritical bifurcation of the limit cycles, i.e., the they coincide on the two-cluster subspace~$C_{(n_1, n_2+n_3)}$.
Then, at $y=0.8$, the three-cluster limit cycle is on the other side of the $C_{(n_1, n_2+n_3)}$ plane. In the transcritical bifurcation, the limit cycles have exchanged their stabilities perpendicular to the two-cluster subspace. Hence, the two-cluster limit cycle is now stable within the three-cluster subspace. 

Analogous to the equivariant transcritical bifurcation of equilibria, in the full system the transcritical bifurcation of limit cycles is an equivariant transcritical bifurcation.
As above, the cluster sizes are not true bifurcation parameters but rather enumerate different invariant subspaces in the same system.
Thus, the transcritical bifurcation takes places simultaneously in all the different subspaces which are related to each other by index permutation, and in the corresponding subspaces which correspond to different cluster sizes~$n_2$ and~$n_3$ at fixed $n_2+n_3=1-n_1$.
In other words, in the full system with~$N$ identical components and sufficiently large~$N$ at $y=0.79625$ multiple limit cycles from different three-cluster subspaces meet in the same two cluster limit cycle, each changing the stability of a different Floquet exponent of the two-cluster limit cycle.  
Since all transversal eigenvalues (i.e., intra-cluster eigenvalues) of the Jacobian matrix are degenerate due to symmetry~\cite{salman2020collective}, it actually affects all of them in the same way. As a consequence, in the full system, the equivariant transcritical bifurcation of two-cluster limit cycles stabilizes the latter and macroscopic two-cluster oscillations should be observable in arbitrarily large systems, i.e. even in the thermodynamic limit. We verified this conclusion numerically for a system with $N = 20$ units and the other parameters as in Fig.~\ref{fig:3d_pp_transcrit}.

%-----------------------------------------------
\section{Discussion and Conclusions}
\label{sec:Conclusions}

In the CO-electrooxidation model we analyzed in this paper, the fully synchronized one-cluster equilibrium loses stability in an $\bbS_N$-equivariant transcritical bifurcation where $N-1$ real eigenvalues change sign simultaneously. Such bifurcations can arise for example as a perturbation of a pitchfork bifurcation~\cite{Golubitsky1988} where the invariant subspace persists. Indeed, the pitchfork bifurcation for identical cluster sizes ($n_1=n_2$) perturbs to a pair of saddle node and transcritical bifurcation for nonidentical cluster sizes. While the symmetric transcritical bifurcation happens simultaneously for all cluster sizes, the parameter values of the saddle node bifurcation are distinct for different relative cluster sizes as seen in Fig.~\ref{fig:2d_map}: The most asymmetric two-cluster state is generated first, then increasingly symmetric ones, and finally the symmetric $(\frac{1}{2}, \frac{1}{2})$-cluster in a pitchfork bifurcation. The symmetry breaking bifurcations from two- to three-cluster equilibria occur in a similar fashion, just that there is an additional cluster as an `bystander'. More generally, if the phase-space geometry allows, $\bbS_N$-equivariant transcritical bifurcation can also have a global flavor~\cite{Ashwin1990}.

Equations that describe globally-coupled identical one-dimensional units also directly relate to more general contexts, such as coupled oscillators. If a general $\bbS_N$-equivariant system undergoes a $\bbS_N$-equivariant transcritical bifurcation, one can describe the dynamics on the corresponding center manifold through a suitable normal form. This approach has been used to describe the dynamics of identical Stuart-Landau oscillators at a symmetry breaking bifurcation\cite{kemeth2019cluster,kemeth20212,fiedler2020global}. 
Truncating to appropriate order, one obtains a cubic $\bbS_N$-equivariant system of coupled one-dimensional variables~$x_k$\cite{elmhirst2001symmetry,Golubitsky2002,stewart2003symmetry,dias2003secondary,kemeth20212, fiedler2020global} that describe the dynamics on the center manifold. In the limit of two large clusters and one cluster of vanishing size, the dynamics is characterized by a web of heteroclinic orbits between the two-cluster equilibria~\cite{fiedler2020global}. These cubic vector fields show the same bifurcation behavior close to the symmetry breaking bifurcations of the system analyzed here, including the transcritical and saddle node bifurcations, as well as the secondary bifurcations where the two-cluster solutions lose stability; cf.~Refs.~\onlinecite{kemeth2019cluster, kemeth20212}. A system of coupled one-dimensional units with a cubic vector field, however, is unable to capture some of the more intricate secondary bifurcation behavior, {such as Hopf and homoclinic bifurcations,} described here. These require the inclusion of higher order terms, such as the coupling of~\eqref{eq:ex} of our CO-electrooxidation model that is beyond cubic order, or a higher-order approximation on the center manifold.

{The dynamical features described in this paper arise in a mathematical model that captures the experimental control of a physical system.
From this perspective, the bifurcations discussed are common to many real-world systems. Yet, we made one simplification which will never be strictly fulfilled in a physical system, namely that all bistable units are identical.
While the permutational symmetry is broken if the units are slightly heterogeneous, hyperbolic equilibria and limit cycles---which are the main focus of our analysis---will persist.
The exact behavior of the branches at the (symmetric) bifurcation points will of course differ as these bifurcations break up into a series of generic bifurcation points; how exactly depends on the heterogeneity. However, the qualitative behavior that is relevant for real-world systems is preserved:
In all ensemble steady states, each of the bistable units will be on one of its three branch segments. 
We can still define a cluster as a group of units that are on the same segment, and a unit transits between the segments at its turning points of $u$, i.e., $u^{\mathrm{p}\to\mathrm{m}}$ or $u^{\mathrm{m}\to\mathrm{a}}$.
However, for the individual units, the turning points do not necessarily occur at identical values of~$u$ as the branches deform.
We verified that Hopf and homoclinic bifurcations still occur in a model of heterogeneous bistable units.
Consequently, our bifurcation analysis facilitates a qualitative understanding of experimental observations also for nearly-symmetric systems and the conditions for the occurrence of oscillations:
Consider, e.g., Li--ion batteries. Here, a global constraint, corresponding to the constant current in our model, is imposed on the total charge. The hysteresis in the cell voltage~$U$ over the total charge which is measured when charging and discharging Li--ion batteries with phase transition cathodes even at infinitely slow rates~[14] could be seen as the `shadow' of the transcritical bifurcations at the turning points of the charge vs voltage curve of the individual nano-sized storage particles in the symmetric system.}

{Another physical system that shows similar symmetry-breaking bifurcations, including a Hopf bifurcation, are certain electronic circuits, which, in addition, can be considered as superlattice models~\cite{Heinrich2010}. While in that case the governing equations of the individual elements were two-dimensional, our studied revealed that stable, macroscopic collective oscillations may also exist in ensembles of coupled one-dimensional bistable units. This might lead to a reinterpretation of the origin of oscillations observed in many-particle systems, where macroscopic oscillations have been interpreted as the result of synchronization of oscillating units.}

%-------------------------------------

\section*{Data Availability Statement}
The data that support the findings of this study are available from the corresponding author upon reasonable request.

%-------------------------------------

\section*{Acknowledgments}

The authors thank A. Bonnefont for input concerning heterogeneous ensembles. CB acknowledges support from the Institute for Advanced Study at the Technical University of Munich through a Hans Fischer fellowship. This research was funded by the Deutsche Forschungsgemeinschaft (DFG, German
Research Foundation) through “e-conversion” Cluster of Excellence
(Grant No. EXC 2089/1-390776260).

%-------------------------------------

\section*{Conflict of interest}

The authors have no conflicts to disclose.

\section*{References}
\bibliographystyle{unsrt}
\bibliography{references}
\end{document}